\documentclass[journal=nalefd,manuscript=communication,layout=twocolumn]{achemso}

\AbstractOn
\usepackage[version=3]{mhchem} 
\usepackage{graphicx}
\usepackage{color}
\usepackage{upgreek}
\usepackage{dcolumn}
\usepackage{bm}
\usepackage[normalem]{ulem}
\usepackage{wasysym} 
\usepackage{amssymb}
\usepackage[mathlines]{lineno}

\usepackage{xcolor}
\usepackage[normalem]{ulem} 
\newcommand\hl{\bgroup\markoverwith
  {\textcolor{yellow}{\rule[-.5ex]{2pt}{2.5ex}}}\ULon}
  
\newcommand\redsout{\bgroup\markoverwith
{\textcolor{red}{\rule[.5ex]{2pt}{0.4pt}}}\ULon}

\newcommand*{\citen}[1]{%
  \begingroup
    \romannumeral-`\x 
    \setcitestyle{numbers}%
    \cite{#1}%
  \endgroup}

\author{Jingyu Duan}	\email{j.duan.17@ucl.ac.uk}
\affiliation{London Centre for Nanotechnology, University College London, London WC1H 0AH, United Kingdom}
\alsoaffiliation{Quantum Motion Technologies, Nexus, Discovery Way, Leeds, LS2 3AA, United Kingdom}
\author{Michael~A.~Fogarty}
\email{m.fogarty@ucl.ac.uk}
\affiliation{London Centre for Nanotechnology, University College London, London WC1H 0AH, United Kingdom}
\alsoaffiliation{Quantum Motion Technologies, Nexus, Discovery Way, Leeds, LS2 3AA, United Kingdom}
\author{James Williams}
\affiliation{London Centre for Nanotechnology, University College London, London WC1H 0AH, United Kingdom}
\author{Louis Hutin}
\affiliation{CEA, LETI, Minatec Campus, F-38054 Grenoble, France}
\author{Maud Vinet}
\affiliation{CEA, LETI, Minatec Campus, F-38054 Grenoble, France}
\author{John J. L. Morton}
\affiliation{London Centre for Nanotechnology, University College London, London WC1H 0AH, United Kingdom}
\alsoaffiliation{Quantum Motion Technologies, Nexus, Discovery Way, Leeds, LS2 3AA, United Kingdom}
\alsoaffiliation
{Dept.\ of Electronic \& Electrical Engineering, UCL, London WC1E 7JE, United Kingdom}

\title{Remote capacitive sensing in two-dimensional quantum-dot arrays}

\date{\today}

\begin{document}


\twocolumn[
    
    \begin{@twocolumnfalse}
        \maketitle
    \end{@twocolumnfalse}
    \vspace{2mm}
    \hrule
    \begin{abstract}
    We investigate gate-induced quantum dots in silicon on insulator nanowire field-effect transistors fabricated using a foundry-compatible fully-depleted silicon-on-insulator (FD-SOI) process. A series of split gates wrapped over the silicon nanowire naturally produces a $2\times n$ bilinear array of quantum dots along a single nanowire. We begin by studying the capacitive coupling of quantum dots within such a 2$\times$2 array, and then show how such couplings can be extended across two parallel silicon nanowires coupled together by shared, electrically isolated, `floating' electrodes. With one quantum dot operating as a single-electron-box sensor, the floating gate serves to enhance the charge sensitivity range, enabling it to detect charge state transitions in a separate silicon nanowire. By comparing measurements from multiple devices we illustrate the impact of the floating gate by quantifying both the charge sensitivity decay as a function of dot-sensor separation and configuration within the dual-nanowire structure.
    {\bf Keywords: Quantum dots, Reflectometry, floating gate coupler, electrostatic coupling}  
    \end{abstract}
    \hrule
    \bigskip
]



Spin qubits in silicon demonstrate the fundamental properties required for scaled quantum computation, with state-of-the-art one- and two-qubit operations demonstrating control fidelities approaching the requirements for fault-tolerant quantum error correction~\cite{yoneda2018a,xue2019,zajac2018, huang2019}. While all control elements have been integrated into single devices with scalable readout mechanisms~\cite{fogarty2018a}, much effort is now being focused into developing these devices from simple laboratory prototype structures into scaled arrays of qubits capable of eventually yielding a quantum advantage~\cite{vandersypen2017a,jones2018logical}.
The promise of a highly developed materials system and mature fabrication industry, together with the success of laboratory, and industry-grade prototype silicon-metal-oxide-semiconductor (SiMOS) quantum-dot based devices~\cite{maurand2016b} has led to the proposition of several approaches to foundry-compatible scaling into grid-based architectures of quantum dot arrays. These approaches range from densely-packed qubits with next-nearest-neighbour couplings~\cite{veldhorst2017a}, dot arrays partially-populated with qubits~\cite{li2018} and arrays with qubit sites linked via mediating structures for remote qubit-qubit coupling~\cite{cai2019}.
	
SiMOS devices which form quantum dots in the corners of silicon nanowires naturally produce bilinear dot arrays~\cite{voisin2014b}, which allow for proximal sensor integration for both charge~\cite{urdampilleta2019, house2016, chanrion2020,ansaloni2020} and spin states~\cite{ciriano-tejel2020}
through dispersive measurements using gate-based reflectometry. The advantages of these integrated sensors can be extended by mechanisms for \textit{off-wire} coupling, to sense the state of dots located in remote locations within the quantum dot array.  
In order to enhance the capacitive coupling between spatially separated quantum dots, studies in planar GaAs/AlGaAs and Ge/Si heterostructures and carbon nanotubes have exploited a \textit{floating gate}~\cite{chan2002,Churchill2009,Hu2007a}; a metallic electrode which is galvanically isolated from, but capacitively coupled to, its immediate environment. 

Here, utilising a single quantum dot sensor, we demonstrate a system capable of performing both proximal and remote capacitive charge sensing within a $2\times 4$ array of quantum dots distributed across two parallel nanowires. We compare these results with geometrically identical single-wire variants, serving as an isolated $2\times 2$ array. Each $2\times$2 array is formed on a single silicon nanowire (SiNW), and all devices described here are located in the same die, fabricated from a fully-depleted silicon-on-insulator (FD-SOI) process~\cite{hutin2016}. Our approach uses floating gate electrodes to capacitively couple a sensor dot to quantum dots on remote nanowires, whilst maintaining sensitivity to adjacent dots within the local nanowire. We quantify the sensitivity to charge movement within these two schemes by experimentally benchmarking the device capacitance matrix, supported by cryo-SiMOS simulations.

\begin{figure*}[]
\centering
\includegraphics[width=1\linewidth]{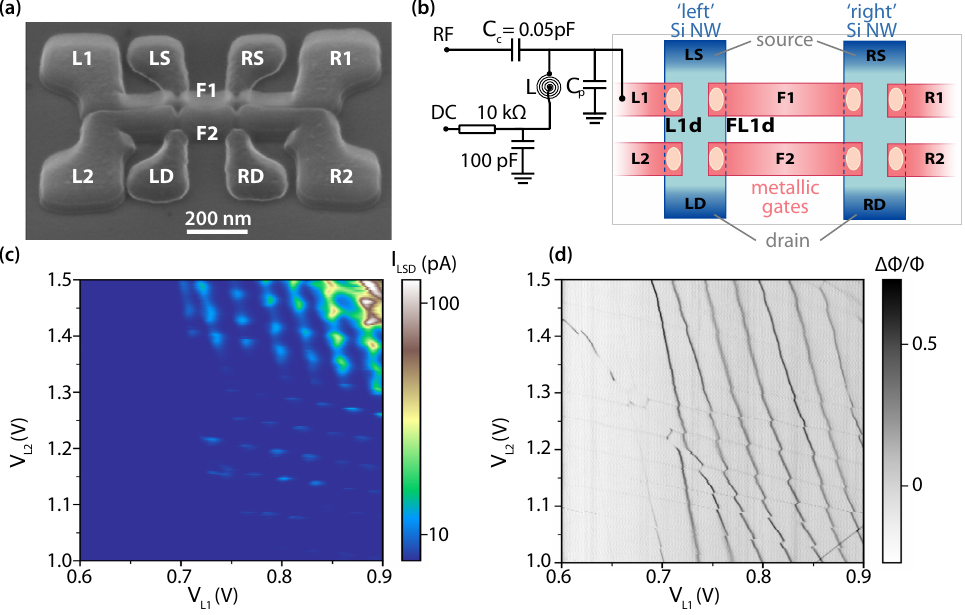}
\caption{Device, measurement and configuration. (a) Oblique-angle Scanning Electron Micrograph illustrating the gate structure of the double-nanowire device coupled by two floating gates. (b) Cartoon of the gate structure and reflectometry circuit diagram. Charge transitions in a variety of quantum dots are detected by the `single electron box' sensor under L1, including those in dots within the same (left, L) nanowire, or located remotely in an adjacent (right, R) nanowire, and detected capacitively through coupling facilitated by the floating gate. (c) Double-dot signatures within the left nanowire through a transport current map of gate L1 vs L2 gate-space with a Source-Drain bias $4$~mV. (d) A concurrent zero-biased reflectometry measurement illustrates dot-lead charge transitions of the L1 sensor dot and capacitive shifts due to addition of electrons to a local quantum dot defined under gate L2.} 
\label{fig:Fig1}
\end{figure*}

The scanning electron micrograph (SEM) image in Fig.~\ref{fig:Fig1}(a) shows a device of the type used in these remote sensing experiments. Two parallel nanowires, with centre-to-centre spacing of 200~nm, are fabricated with two central floating gates F1 and F2 wrapping the interior edges of both, spanning the gap between the two silicon structures. Gates F1 and F2 are capacitively coupled to the surrounding gates by proximity, but are otherwise electrically isolated. All gate structures are separated by a SiN spacer which increases cross capacitance. The device is further encapsulated by 300~nm of silicon oxide, above which an additional top gate T is deposited utilising a back-end metallisation layer (not shown). Full geometric details for the family of devices compared in this work can be found in Supplementary \S I. 
The charge sensor for these experiments consists of a two-terminal structure in which a charge island is connected to single reservoir, known as a single electron box (SEB)~\cite{urdampilleta2019, house2016,chanrion2020,ansaloni2020, ciriano-tejel2020}. The sensor is configured under a single gate, L1, utilising the dot L1d, which is coupled to an electron reservoir and measured using the reflectometry circuit depicted in Fig.~\ref{fig:Fig1}(b). With this configuration, the addition of electrons to the dots within the left nanowire can be inferred from either the transport current $I_{\rm SD,L}$ through the device with source-drain bias $V_{\rm SD}=4$~mV, seen in Fig.~\ref{fig:Fig1}(c), or the S$_{11}$ reflectometry signal $\Delta\Phi/\Phi$ (measured at $V_{\rm SD}=0$~V) seen in Fig.~\ref{fig:Fig1}(d), which maps the same gate voltage space. Both measurements contain structure attributed to multiple dots within the 2$\times$2 array of the left nanowire. Due to the low transport current through the device, discerning the occupancy of the dots via transport is a significant challenge, while the capacitive shifts due to the addition of an electron are readily detected in reflectometry, which can probe all proximal quantum dots down to the last electron transition (Supplementary \S I). The SEB dot-lead transitions at lower SEB electron numbers are less visible due to the reduction in tunneling rates below  the RF frequency of the reflectometry measurement~\cite{Gonzalez-Zalba2015}.

\begin{figure*}[]
\centering
\includegraphics[width=1\linewidth]{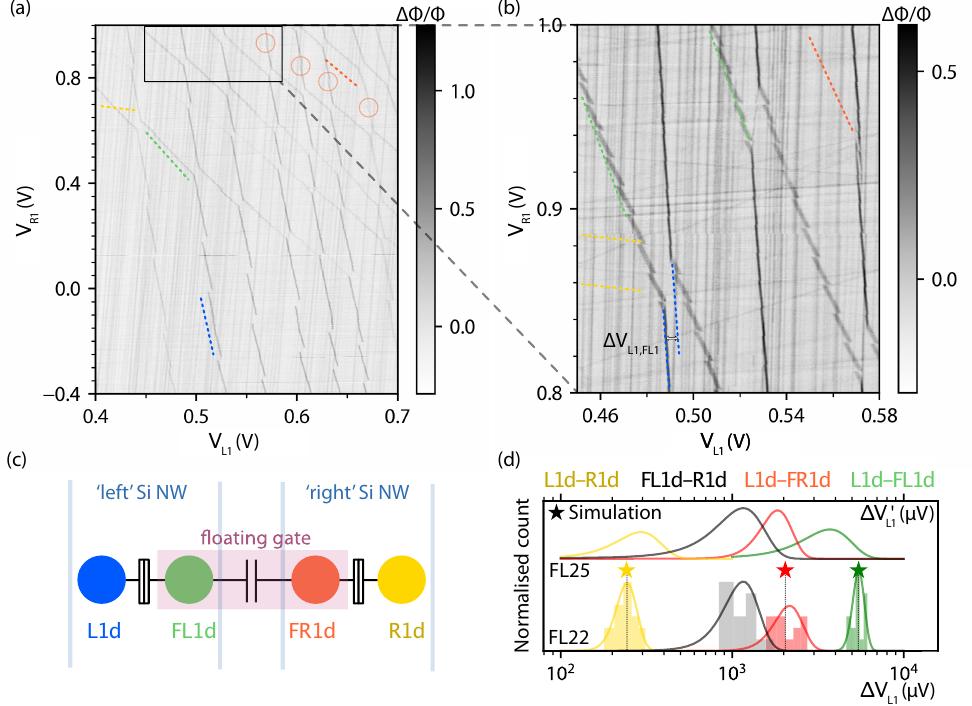}
\caption{Remote sensing of quantum dot charge transitions in different silicon nanowires using the floating gate. (a) Charge stability map in the L1 and L2 gate-space (top gate T potential $V_{\rm T} = 4$V; $V_{\rm L2}$ and $V_{\rm R2}$ = $-1$V). (b) Zoom-in of (a) illustrating the different capacitive shifts of the sensor dot-lead transition due to loading electrons into different quantum dots along the floating gate direction. (c) Schematic of the remote sensing showing quantum dots as a network of charge nodes and capacitors. Dot-lead transition in stability diagram are indicated by dashed line with corresponding quantum dot color. (d) Histogram of capacitive shifts induced on the sensor dot by a charge transition in another quantum dot measured at various anticrossing, following the colour-coding in (c), normalized as a dot L1 Coulomb peak shift $\Delta V$. The coloured solid-line is the normalized fit to a Gaussian probability density function of histogram which attribute to the same quantum dot. ($\bigstar$) show calculated values from a COMSOL finite element simulation. Grey curves and histogram represent capacitive shifts from transitions in dot R1d measured using FL1d dot-lead transition. All data described above are from device `FL22' --- a normalized fit of the $\Delta V'$ histogram from a similar device `FL25' is shown vertically offset above.} 
\label{fig:Fig2}
\end{figure*}

As the floating gates are galvanically isolated, we use the top metal gate T to assist in the accumulation of quantum dots under floating gates, primarily via the mutual capacitance between gates F1, F2 and T (see Supplementary \S V). Simultaneously, both $V_{\rm L2}$ and $V_{\rm R2}$ are set to a depletion mode to avoid formation of quantum dots under gates L2, R2 and F2, to effectively `shut-off' the lower half of the device by electron depletion. With the voltage sweep of $V_{\rm L1}$ and $V_{\rm R1}$ shown in Fig.~\ref{fig:Fig2}(a), and noting the influence of the floating gate F1 which is capacitively coupled to both active gates, we can load electrons into dots L1d and R1d, as well as dots FL1d and FR1d, from their neighbouring reservoirs. Charge detection of these four distinct quantum dots is shown in the stability diagram measured in the reflectometry phase signal Fig.~\ref{fig:Fig2}(a), and includes the remote sensing of dots FR1d and R1d, located in the `right' SiNW, detected by the sensor dot L1d, located in the `left' SiNW. The sensor dot L1d is estimated to hold $\approx$10 electrons in this voltage range, where dot-reservoir charge transitions can be observed directly as a phase peak. We can then identify the remaining three different quantum dots capacitively coupled to the sensor through two complementary approaches:
\begin{itemize}
    \item[(1)] Through the ratio of cross capacitance between the two active gate voltages $V_{\rm L1}$ and $V_{\rm R1}$ and the dot.
    \item[(2)] Through direct charge detection by the sensor dot, assessing the magnitude of the capacitive shift upon the sensor.
\end{itemize}
For the voltage map between $V_{\rm L1}$ and $V_{\rm R1}$ shown in Fig.~\ref{fig:Fig2}(a), each of the four dots capacitively couple to the L1 and R1 electrodes with differing strength, and we illustrate the four quantum dots present with reference to the colour code shown in the capacitance connectivity diagram of Fig.~\ref{fig:Fig2}(c). In Figs.~\ref{fig:Fig2}(a) and \ref{fig:Fig2}(b), the blue dashed line indicates dot-lead charge transition of the SEB, L1, which naturally has the highest lever arm to $V_{\rm L1}$. The other three coloured dashed lines highlight each remaining variety of dot-lead charge transition. The floating-gate-induced quantum dot in the left SiNW FL1d (green) is more strongly coupled to the sensor gate L1 due to its proximity, while in the SiNW on the right, the other floating-gate-induced quantum dot FR1d (red) and gate-induced quantum dot R1d (yellow) are more strongly coupled to gate R1. When quantum dot FL1d is sufficiently occupied, the increase in tunnel rates allows for FL1d dot-lead transitions to also be directly detected in the reflectometry phase shift signal. This signal allows us to trace back the number of electrons in sensor dot L1d (see Supplementary \S I). This approach can be further quantified by comparing the cross-capacitance ratios $\alpha_{(i,j)}$ calculated as the degree to which gate L1 influences the other dot-lead transitions in voltage space. Assuming $\alpha_{\rm (L1,L1)}=1$, this method yields $\alpha_{\rm (FL1,L1)} = 0.173$, $\alpha_{\rm (FR1,L1)}=0.124$, $\alpha_{\rm (R1,L1)}=0.005$. A significant drop in the cross-capacitance ratio is therefore apparent for groups of dots under spatially separated gates.

A second quantitative approach to distinguish the different quantum dots coupled to the sensor is to analyse the strength of the capacitive coupling between the sensor dot L1d and each of the remaining dots.
In Fig.~\ref{fig:Fig2}(d) we plot a histogram of the shifts $\Delta V_{({\rm L1},i)}$, expressed in terms of the gate L1 voltage $V_{L1}$, arising from the capacitive shift in the sensor dot L1d due to the addition of an electron to some other dot $i$~\cite{vanderwiel2003a}. We use a peak-finding algorithm near a capacitive shift of interest in Fig.~\ref{fig:Fig2}(b) and take the difference between the shifted dot-lead reflectometry peaks, extrapolated to the same value of $V_{\rm R1}$. The capacitive shifts extracted in this way group naturally into three distinct sets, each corresponding to the transitions in another quantum dot indicated following the colour code in Fig.~\ref{fig:Fig2}c). 
Being located in the same nanowire, FL1d (green) is the most strongly coupled to the sensor dot, while the other floating-gate-induced quantum dot FR1d (red), located in the remote nanowire, shows a slightly weaker coupling. The R1 gate-induced quantum dot R1d (yellow) in the remote nanowire shows the weakest coupling, but can still be detected.
A normalized fit of the probability density function of each group provides the mean capacitive shift referenced against the sensor dot gate voltage: $\Delta \overline{V}_{\rm (L1,FL1d)}$ = 5.47 mV, $\Delta \overline{V}_{\rm (L1,FR1d)}$ = 2.16mV, $\Delta \overline{V}_{\rm (L1,R1d)}$ = 0.243 mV. These values show good agreement to simulations of the capacitance matrix for this device structure (see Supplementary \S III). 

As certain charge transitions FL1d are directly visible in the phase response, we can also extract a corresponding capacitive shift between dots FL1d and R1d, which is the symmetric analogue to the sensor dot coupling through the floating gate to FR2d.
Data corresponding to such $\Delta V_{\rm (L1d,FR1d)}$ shifts are shown in grey in Fig.~\ref{fig:Fig2}d), and indeed fall within a similar range to $\Delta V_{\rm (L1,FR1d)}$. This asymmetry is not captured in our simulations and is most likely due to finite lithographic misalignment between the patterns of the nanowire and the split between the gates. Based on automated overlay controls and tools specifications we estimate that the cuts, although centered on the nanowires by design, are probably shifted by 5-10nm on a typical device. In our case, this asymmetry translates into stronger lever-arm parameters for the dots defined along the right edges of the nanowires, and is systematically observed in other devices~\cite{ciriano-tejel2020,Ibberson2020}. 
Finally, to show the consistency of these values across different devices of the fabricated on the same die, we performed the same set of measurements on a second device and plot the extracted Gaussian fits to $\Delta V_{(i,j)}$ for each pair of dots on the same axis in Fig.~\ref{fig:Fig2}(d).

\begin{figure}[]
\centering
\includegraphics[width=1\linewidth]{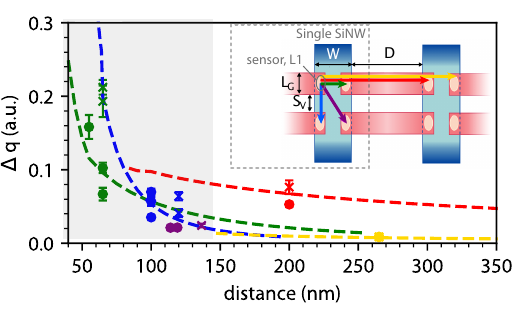}
\caption{Normalized capacitive coupling as a function of distance. Voltage shifts in the sensor dot arising from capacitive coupling to other quantum dots are normalized against the addition voltage of the individual SEB to compare measurements from two floating gate devices and three single-nanowire devices ($\times$, $L_{\rm g}=60$~nm; $\CIRCLE$, $L_{\rm g}=50$~nm). Arrows in the inset illustrate the type of sensing: green, blue and purple data points relating to sensing within a single-nanowire and are obtained from both type of devices. Red and yellow data points
required floating gate devices). COMSOL simulations are used to obtain parameter sweeps relating to each class of dot being sensed, following the colouring in the inset --- a single normalisation is applied to all simulated curves. Error bars in the data include the uncertainties in both the capacitive voltage shifts and addition voltages.} 
\label{fig:Fig3}
\end{figure}
To demonstrate the enhancement of capacitive coupling arising from the floating gates, we compare results from floating gate devices with those from devices with similar dimensions containing only single, isolated silicon nanowires (see Supplementary \S I). In order to facilitate the comparison of results from different devices, sensor dots and lever arms, we adopt a measure of the SEB sensitivity to the charge transitions in nearby quantum dots based on normalizing the voltage-referenced capacitive coupling by the addition voltage required to add an electron to the SEB: 
$\Delta q= \Delta V_{({\rm L1},i)}/V_{\rm C_{L1d}}$, where $\Delta V_{({\rm L1},i)}$ is the detected voltage shift in $V_{\rm L1}$ arising from coupling to dot $i$, $V_{\rm C_{L1d}}$ is the change in $V_{\rm L1}$ required to add an electron to the sensor dot L1d.

We first study the normalized SEB charge sensitivity within a 2$\times$2 quantum dot array located in a single silicon nanowire. Here we can compare the capacitive coupling between dots formed on opposite edges of the nanowire, between adjacent dots formed along a common SiNW edge, and also between diagonally coupled, next-nearest neighbour quantum dots. These configurations are shown in the inset of Fig.~\ref{fig:Fig3}, which also illustrates configurations for sensing dots in a neighbouring SiNW, with coupling facilitated through the floating gate. The data in Fig.~\ref{fig:Fig3} compare three single-nanowire devices, each consisting a 2$\times$2 quantum dot array, as well as the corresponding single-nanowire arrays within two floating-gate devices. The \emph{intra-wire} normalised sensitivities $\Delta q$ fall off quickly with increasing separation between the quantum dots, though a single power law cannot be used to describe the overall trend with distance for all couplings, due to the difference in mutual capacitance for dots located on the same or opposite edges of the nanowire. 

Modelling the quantum dot as conducting ellipsoids we calculate the Maxwell capacitance matrix for varying center-to-center dot separation $d$, along with other nanowire design parameters (full details of the simulation method and relation to device dimensions are given in Supplementary \S I--V).
Each of the parametric sweeps from the simulations (dashed lines in Fig.~\ref{fig:Fig3}) settles to a power law attributed to each sensor-dot configuration: nearest-neighbour couplings along the edge of the nanowire (L1d-L2d) or across the nanowire (L1d-FL1d) have couplings which decay approximately as $\Delta q\propto d^{-2.8}$ or $d^{-2.5}$ respectively, over the range of distances studied here. 
Data for a next-nearest neighbour configuration L1d-FL2d, where the dots are positioned diagonally across the wire, is shown for completeness but is not modelled.
For the `remote sensing' configuration where charge transitions are detected through the floating gate, the normalized capacitive coupling is sustained over a much greater distance, as reflected in the experimental data and simulations. By sweeping the floating gate length (approximated to be the SiNW separation, $D$) simulations show that the two dots under each corner of the floating gate have a coupling which is dominated by the second-order capacitive coupling via the floating gate at these distances, and decays only as $\propto D^{-0.4}$ (see Supplementary \S VI). Combined with the additional spacing of the nanowire width local to the SEB, this results in an coupling decay for the dot L1d-FR1d configuration which can be approximated as $\Delta q \propto d^{-0.6}$ in the range studied here. As a result, the mutual capacitive shift for dot L1d-FR1d remains relatively high, even at distances exceeding 300~nm, as shown in Fig.~\ref{fig:Fig3}. 

Coupling the sensor to dot R1d now involves three degrees of separation from the sensor, with a corresponding drop in sensitivity for short separations. However, the action of the floating gate leads to a much more gradual decay in sensitivity with distance that goes as $\Delta q \propto d^{-0.7}$ in our simulations. As a result, for distances above $d\approx220$~nm, the floating gate mediated coupling between dots arranged on opposite edges of \emph{different} nanowires exceeds that from two dots on opposite edges of the same silicon nanowire (see Supplementary \S VI).
Furthermore, the charge distribution due to floating gate geometry could be optimised to yield a stronger absolute coupling, while maintaining the much more gradual decay with distance
~\cite{trifunovic2012a}.


Our experimental measurements and simulations indicate decays in capacitive coupling strength which fall off more slowly than $\propto d^{-3}$, as previously observed within arrays of Si/SiGe planar quantum dots~\cite{Zajac2016,neyens2019}. However, such measurements were made within planar quantum dot devices with a high density of metallic gate electrodes, expected to screen mutual capacitive coupling. Indeed, considering only the first-order approximations to capacitive couplings, our simulations also show decays that approach $d^{-3}$ (see Supplementary \S VI). 
In contrast, the devices studied here contained a relatively low density of metallic gate electrodes, and the fabrication of the split-gates involved etching of metal that was replaced by SiN. The result is a reduced decay rate in sensitivity as a function of dot-dot separation --- most strikingly when facilitated by the capacitively coupled floating gate. Instead of screening charge movement, the floating gate propagates the effect of charge movement over a distance to be chosen as a design parameter, coupling charge between two otherwise separate silicon structures. While the simulations are able to capture well the trends in the different classes of coupling, the residual spread in experimental values across the measurements may be due to the asymmetry in realistic devices, not captured by the simulations, which can influence not only the dot to dot geometrical distance but also the device lever arms.

The capacitive shifts we measured between QDs, both locally and on distinct nanowires, are well above the full-width at half-maximum (FWHM) of the SEB dot-lead charge transition (see Supplementary \S II). Assuming a Lorentzian lineshape for the measured SEB charge transition, any capacitive shift greater than twice the FWHM gives at least $94\%$ of the maximum sensor contrast (e.g.\ for spin-dependent tunneling readout). Based on our simulations and the intrinsic FWHM of the sensor transition of 0.24~mV, dot L1d-FR1d type couplings mediated by the floating gate could be used to achieve spin readout for distances up to 500~nm without a reduction in readout contrast.

In addition to applications for sensing, capacitive coupling has been used to realise local multi-qubit interactions in a variety of systems, including singlet-triplet qubits~\cite{shulman2012} and charge qubits~\cite{li2015a,macquarrie2020}. Meanwhile, several approaches to scaling quantum dot arrays pursue long-range coupling between qubits to facilitate the integration and fan-out of control electronics and suppress charge leakage~\cite{vandersypen2017a,cai2019}--- solutions to realising such two-qubit gates include exploiting a RKKY mediating exchange interaction~\cite{malinowski2019,cai2019} or coupling via a superconducting resonator~\cite{borjans2020}. Multi-qubit operations utilising capacitive coupling via floating gates, coupling two singly-occupied planar dot structures, have been proposed to produce a spin-spin coupling Hamiltonian $H_{\rm S-S} \simeq J_{12}(\sigma_x^1\sigma_x^2+\sigma_y^1\sigma_y^2)$ when the Zeeman energy $E_{\rm Z} \gg J_{12}$ and where $\sigma_{x,y,z}$ are the Pauli matrices in the relevant qubit basis~\cite{trifunovic2012a}, which can be used to implement the iSWAP operation~\cite{schuch2003}. Combining the assumptions within Ref~[\citen{trifunovic2012a}] with the parameters of the devices studied here and spin-orbit coupling strength for silicon~\cite{tanttu2019}, we estimate a coupling of $H_{\rm S-S} \simeq 10^3$~Hz under realistic device operating conditions between FL1d and FR1d with nanowire separation $\sim200$nm, which is too weak for practical applications. However, utilising the floating gate to couple two singlet-triplet qubits via $H_{\rm ST-ST} \simeq J_{12}/2((\sigma_z-I)\otimes(\sigma_z-I))$~\cite{shulman2012}, where $I$ is the identity matrix, exploits the much stronger electric-dipole coupling to achieve the CZ operation. For the nanowire geometry presented here (i.e.\ with singlet-triplet qubits arranged on each nanowire and the nearest dots of each pair separated by $\sim200$nm) we estimate $H_{\rm ST-ST}\simeq 10^{12}$~Hz via the model in Ref.~[\citen{trifunovic2012a}], made more favourable in this geometry due to reduced oxide thickness. In SiGe devices, coupling between charge qubits $H_{\rm C-C} \simeq g/4((I-\sigma_z)\otimes(I-\sigma_z))$ mediated by the mutual capacitance term~\cite{li2015a,macquarrie2020} has been demonstrated with a strength of $\approx$15~GHz over dot separations of 130~nm~\cite{macquarrie2020}, while for the device geometry studied here our results predict $H_{\rm C-C}\simeq 10^{11}$~Hz for dots separated by 200~nm on different nanowires.

We have demonstrated through experiments and simulation the effect of integrating floating gate electrodes to extend the sensitivity range of a single capacitive sensor, highlighting in particular the potential to couple quantum dots located on distinct silicon nanowires. Our measurements made the use of a single electron box charge sensor, while we note that a parallel study on similar devices 
illustrates an alternative mode for charge detection in such structures, with one nanowire acting as a single electron transistor that remotely senses the charge occupancy of dots on the other nanowire~\cite{gilbert2020}. In future devices with overlapping gate architecture~\cite{zajac2018}, a second layer of gate electrodes could be used independently tune the quantum dots confined under the floating gates and achieve remote interactions. Given the substantial promise of spin qubits formed along quasi-1D arrays, along the edges of silicon nanowires~\cite{ciriano-tejel2020}, the enhanced capacitive couplings we measure using floating gates provide a potential route to couple qubits distributed across separate nanowires and thus for scaling in a second dimension. \\

\begin{acknowledgement}
The authors gratefully acknowledge the financial support from the European Union's Horizon 2020 research and innovation programme under grant agreement No 688539 (http://mos-quito.eu); as well as the Engineering and Physical Sciences Research Council (EPSRC) through the Centre for Doctoral Training in Delivering Quantum Technologies (EP/L015242/1), QUES$^2$T (EP/N015118/1)
and the Hub in Quantum Computing and Simulation (EP/T001062/1).
\end{acknowledgement}

\providecommand{\latin}[1]{#1}
\makeatletter
\providecommand{\doi}
  {\begingroup\let\do\@makeother\dospecials
  \catcode`\{=1 \catcode`\}=2 \doi@aux}
\providecommand{\doi@aux}[1]{\endgroup\texttt{#1}}
\makeatother
\providecommand*\mcitethebibliography{\thebibliography}
\csname @ifundefined\endcsname{endmcitethebibliography}
  {\let\endmcitethebibliography\endthebibliography}{}

\end{document}


\maketitle
\section{Additional Device Data}\label{SupSec:AdditionalDevices}
\begin{figure*}[ht]
\centering
\includegraphics[width=1\linewidth]{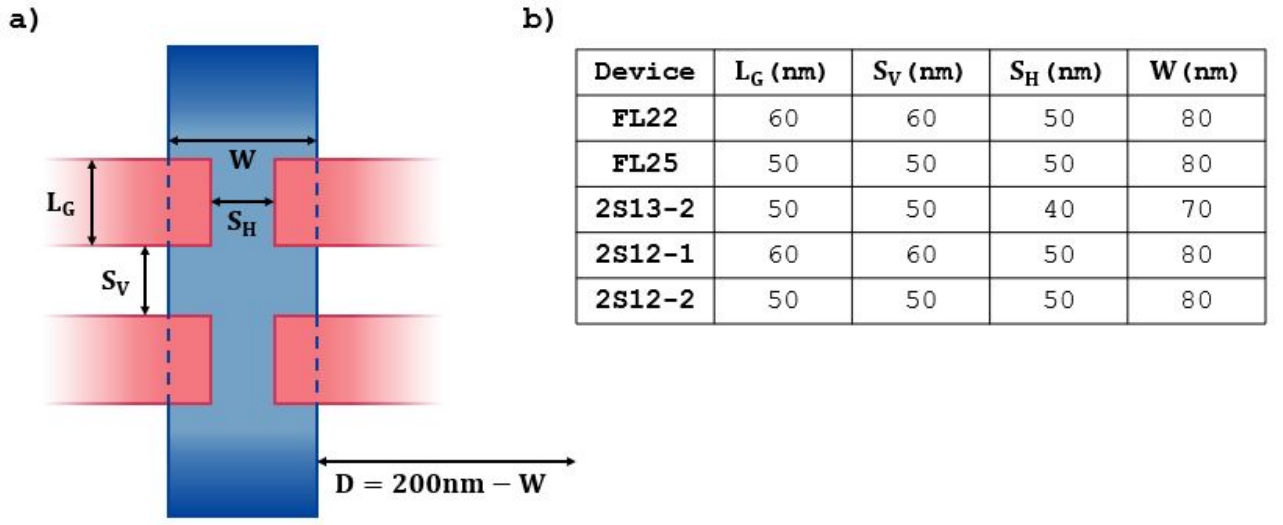}
\caption{(a) Single nanowire with relevant dimensions indicated. (b) Tabulated dimensions for the different device used in this work} 
\label{SupFig:FigS_DeviceComparison}
\end{figure*}

The results from five devices are compiled to produce the data illustrated in Fig. 3 in the main text. Each device is configured with the same reflectometry measurement setup as illustrated in Fig. 1(b). The core dimensions associated with the devices are depicted in Fig.~\ref{SupFig:FigS_DeviceComparison}a), where devices with floating gates are identical replications of the same 2$\times$2 bilinear array, with separation distance of $D = 200$~nm$-W$ as measured edge-to-edge from the nanowires, where $W$ is the total nanowire width.   

The core dimensions of the five different devices used to compile the data in the main text are tabulated in Fig.~\ref{SupFig:FigS_DeviceComparison}b). Devices which couple nanowires via floating gates are identified by the `FL' key, while the remaining devices are simple 2$\times$2 arrays formed in a single nanowire.

\begin{figure*}[ht]
\centering
\includegraphics[width=1\linewidth]{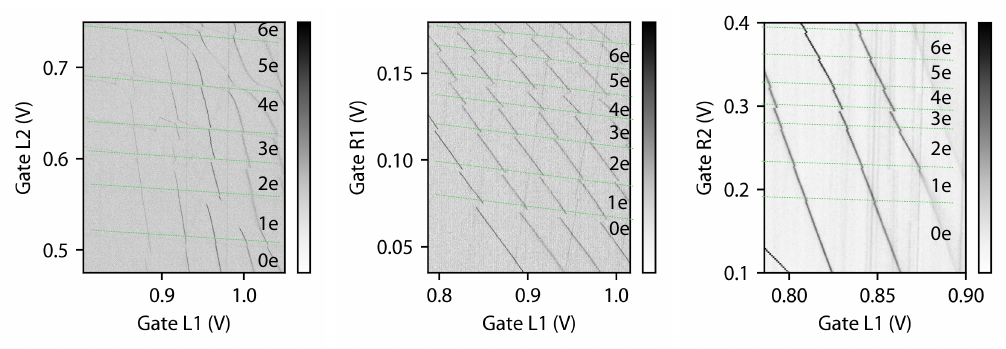}
\caption{Example charge stability diagram (Device ID 2S13-2), showing all three dot-sensor contributions in the $2\times2$ array, individually operated in a double-dot regime} 
\label{SupFig:FigS_2S13-2}
\end{figure*}

Example measurements from a 2$\times$2 array under an single nanowire are illustrated in the stability diagrams in Fig.~\ref{SupFig:FigS_2S13-2}, showing double-dot behavior for pairwise combinations of the four voltage control inputs, demonstrating controlability over the four dot locations. The stability diagrams are achieved by sweeping the relevant control inputs, while holding the remaining voltages low enough to deplete the remaining dots of electrons.

Fig.~\ref{SupFig:FigS_2S13-2} shows, from left to right, loading the first six electrons to dots $D_{L2}$, $D_{R1}$ and $D_{R2}$ in the `2S13-2' device, with dimensions listed in Fig.~\ref{SupFig:FigS_DeviceComparison}b). The capacitive shift of the sensor dot $D_{L1}$ probes charge transitions in the other three quantum dots. An even stronger capacitive shift due to an intentional ion-implanted donor appears in some of the stability diagrams. 

\newpage
\section{Sensor power broadening}\label{SupSec:SensorPower}
\begin{figure*}[ht]
\centering
\includegraphics[width=0.8\linewidth]{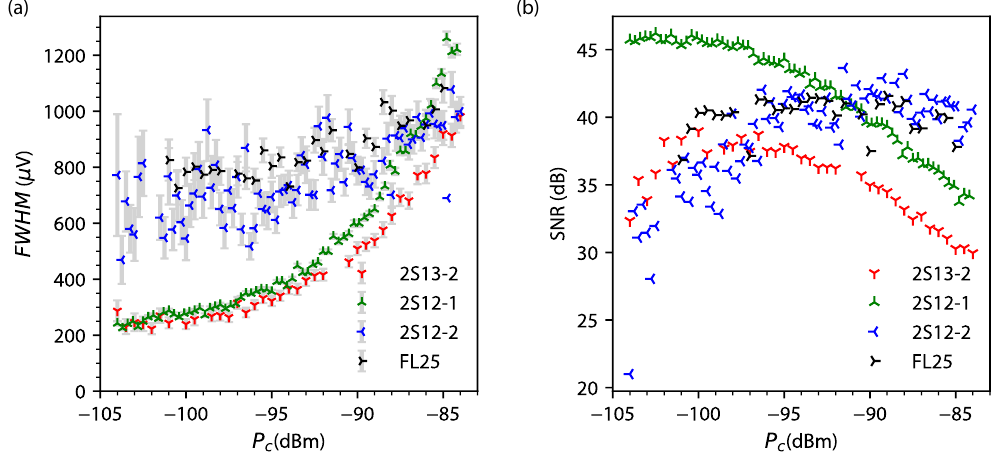}
\caption{(a) sensor dot-lead signal linewidth full width half maximum (FWHM) as a function of power delivered at device. (b) signal to noise ratio as a function of power delivered at device with integration time 0.4 ms.} 
\label{SupFig:FigS_power}
\end{figure*}

We measure the RF-power dependence of different sensor dot-lead Coulomb peaks. We vary the input carrier power $P_{\rm c}$ at resonance frequency to observe the broadening of sensor linewidth $FWHM$ and its impact on the signal-to-noise ratio (SNR). SNR is defined here as $S^{2}/\sigma^{2}_{bg}$ where $S$ is the magnitude of the reflectometry Coulomb peak and  $\sigma_{\rm bg}$ is the standard deviation of the background noise. The large values of SNR recorded are a result of long integration times (0.4~ms).  

As shown in Fig.~\ref{SupFig:FigS_power}, all four dot-lead linewidths reach intrinsic limits (due to dot-lead tunnel rates and near neighbour reservoir electron temperature) for $P_{c}  <  -90$~dBm. The selected dot-lead transition of `FL25' and `2S12-2' experience much higher tunnel rates which in term lead to lower SNR. When $P_{c} > -90$~dBm, the dot-lead linewidths are dominated by the effects of the RF drive used the reflectometry measurement, with a corresponding degradation in SNR. 

\newpage
\section{Cryo-MOS Simulation Details}\label{SupSec:CryoMOS}
The nanowire devices were modelled using the geometrical information given from the fabrication dies. The electrostatics, semiconductor and Schr\"odinger-Poisson modules were used within COMSOL to estimate the quantum dot size, electrostatics, capacitance matrices and lever arms presented and used throughout this work. 

The Maxwell capacitance matrix is used to compute expected values of $\Delta q$ for the sensor dot L1d~\cite{vanderwiel2003a} to enable direct comparison with the measured values. The quantum dots were modelled as conducting ellipsoids closely matching the asymmetric manifolds defined by a Schr{\"o}dinger-Poisson study of the single electron effective mass approximation under the device geometry, with the dependence of the mutual capacitance as a function of separation for each sensor-dot configuration given by the center-to-center distance, $d$,  between the overlap regions of the nanowire(s) and gate electrodes.

\begin{figure*}[ht]
\centering
\includegraphics[width=0.8\linewidth]{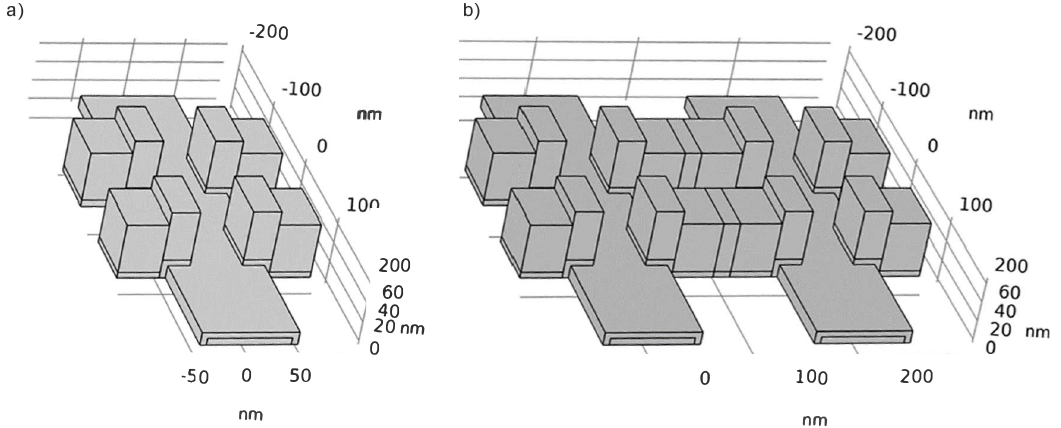}
\caption{(a) Single nanowire $2\times2$ array and (b) two parallel nanowires with floating gates, producing a $2\times4$ array of dots} 
\label{SupFig:FigS1_DeviceCompare}
\end{figure*}

Figure \ref{SupFig:FigS1_DeviceCompare} illustrates a single- and double-nanowire device of identical dimensions constructed in COMSOL. The top gate metallic electrode $G_{\rm T}$ is not shown. The back-gate, source and drain are held constant for all experiments, and are presumed to only add small indirect coupling within the elements of the device and are accordingly not modelled. 

To estimate the dot size, a Schr\"odinger and Poisson study is undertaken by transforming the interface above the silicon into a boundary condition characterised by the material properties including oxide thickness and material work functions. The gate inputs were taken from experimental data, with the dot size resulting from the outputted probability density ($\rho(x,y,z)=|\psi(x,y,z)|^2$), based on the calculate conduction band from the device simulated a $T=2$~K, with the assumption that the electrons are fully confined inside the silicon nanowire. The quantum dot was then modelled as a perfect conductor ($\epsilon_{\rm r}=1$) within the device for use in the electrostatic simulation. 

The collected dots and gates were then swept within an electrostatic study to find the full range of capacitance measurements needed to find the Maxwell capacitance matrix, for a model of the silicon nanowire device. The lever arms for each terminal could then be measured using $\alpha_{(i,j)}=-C_{(i,j)}/C_{(j,j)}$, with $C_{(j,j)}$ and $C_{(i,j)}$ being the self and mutual capacitance respectively.

Each parameter sweep described in the main text was generated as a symmetric device about the centre of the floating gate. For the case of sweeping a variable such as $S_{\rm v}$, $S_{\rm h}$, or the nanowire to nanowire separation $D$, with the dot size and position fixed in relation to the area of gate overlap on the silicon nanowire, ensuring that the dot position matches the change in position of the gate as a parameter is swept.

As the dots defined under the floating gates have stronger dependence on other voltage input parameters, such as $V_{\rm T}$ (see Supplementary \S\ref{SupSec:2x4CapMat} for full discussion) it is possible that the dots under the floating gate have a different distribution of the electron wavefunction than dots induced under a gate connected directly to an input control voltage, indicating a free parameter in our simulations. As the voltage applied to the floating gate can only be \emph{estimated} from simulations, while the mutual capacitance can be physically \textit{measured}, we must account for this missing voltage setting (which dictates dot size and location) in the simulations by matching the observable results shown in Fig. 2(d). Our capacitance model treats the dot wavefunction as a solid ellipsoid with a boundary set to capture 70\% of the electron probability density generated from the Schr\"odinger and Poisson study discussed above.

In order to account for the unknown electrostatic effect of the floating gate, the size of the `inner' dots under this gate (FL1d, FL2d, FR1d, FR2d) was scaled uniformly as a free parameter, with the best fit to the data being 40$\%$ of the size of the `outer' dots (L1d, L2d, R1d, R2d). The simulated mutual capacitance was calculated as a function of this free parameter and was compared to the experimental data. Finally, the ratios of mutual capacitance between L1d to FL1d, FR1d, R1d were optimised to match experimental data by applying minor perturbations ($\mathcal{O}\approx1$)~nm to the position of the `outer' dots.

\newpage
\section{Single-nanowire 2$\times$2 array capacitance and lever arm matrix}\label{SupSec:2x2CapMat}
From the simulations of the device shown in Fig.~\ref{SupFig:FigS1_DeviceCompare}(a), the capacitance matrix is extracted and tabulated in Table~\ref{SS_Tab:2x2CapMat}.

\begin{table}[ht]
  \centering
    \begin{tabular}{|c||c|c|c|c|c|c|c|c|c|}
    \hline
      & L1 & L2 & R1 & R2 & L1d & L2d & R1d & R2d  & T\\
  \hhline{|=||=|=|=|=|=|=|=|=|=|}
    L1&41.3330 & -9.1352 & -8.9286 & -1.7809 & -5.5588 & -0.1328 & -0.4838 & -0.0827 & -6.1058 \\
    \hline
    L2&-9.1351 & 41.3180 & -1.7800 & -8.9219 & -0.1322 & -5.5537 & -0.0827 & -0.4838 & -6.1044 \\
    \hline
    R1&-8.9286 & -1.7799 & 41.3500 & -9.1300 & -0.4852 & -0.0830 & -5.5497 & -0.1319 & -6.1183 \\
    \hline
    R2&-1.7810 & -8.9218 & -9.1300 & 41.3330 & -0.0828 & -0.4848 & -0.1320 & -5.5429 & -6.1162  \\
    \hline
    L1d&-5.5587 & -0.1323 & -0.4853 & -0.0827 & 6.7578 & -0.0090 & -0.0999 & -0.0071 & -0.0715  \\
    \hline
    L2d&-0.1329 & -5.5535 & -0.0829 & -0.4848 & -0.0090 & 6.7524 & -0.0071 & -0.0998 & -0.0713 \\
    \hline
    R1d& -0.4838 & -0.0826 & -5.5495 & -0.1321 & -0.0999 & -0.0071 & 6.7459 & -0.0089 & -0.0715 \\
    \hline
    R2d&-0.0826 & -0.4838 & -0.1320 & -5.5428 & -0.0070 & -0.0998 & -0.0089 & 6.7389 & -0.0714 \\
    \hline
    T&-6.1055 & -6.1050 & -6.1173 & -6.1165 & -0.0715 & -0.0714 & -0.0713 & -0.0714 & 94.8570 \\
    \hline
    \end{tabular}%
    \caption{Maxwell capacitance matrix (in aF) for the single-nanowire $2\times2$ array of quantum dots.}
\label{SS_Tab:2x2CapMat}
\end{table}%

The lever arm matrix which transforms voltage space from our vector of voltage control inputs $\bm{V}$, to the vector of chemical potential on each dot $\bm{\mu}$ takes the form $\bm{\mu = e\alpha V}$, where elements of $\bm{\alpha}$ are $\alpha_{(i,j)} = -C_{(i,j)}/C_{(j,j)}$, and can be drawn from the values in Table~\ref{SS_Tab:2x2CapMat}:

\begin{equation}
\begin{pmatrix}
\mu_{\rm L1d}\\
\mu_{\rm L2d}\\
\mu_{\rm R1d}\\
\mu_{\rm R2d}\\
\end{pmatrix}
= e\times 10^{-2}
\begin{bmatrix}
13.46 & 0.32 & 1.17 & 0.20 \\
0.32 & 13.44 & 0.20 & 1.17 \\
1.17 & 0.20 & 13.41 & 0.32 \\
0.20 & 1.17 & 0.32 & 13.41 \\
\end{bmatrix}
\begin{pmatrix}
V_{\rm L1}\\
V_{\rm L2}\\
V_{\rm R1}\\
V_{\rm R2}\\
\end{pmatrix}\label{SS_Eq:2x2LA}
\end{equation}

\begin{figure*}[ht]
\centering
\includegraphics[width=0.6\linewidth]{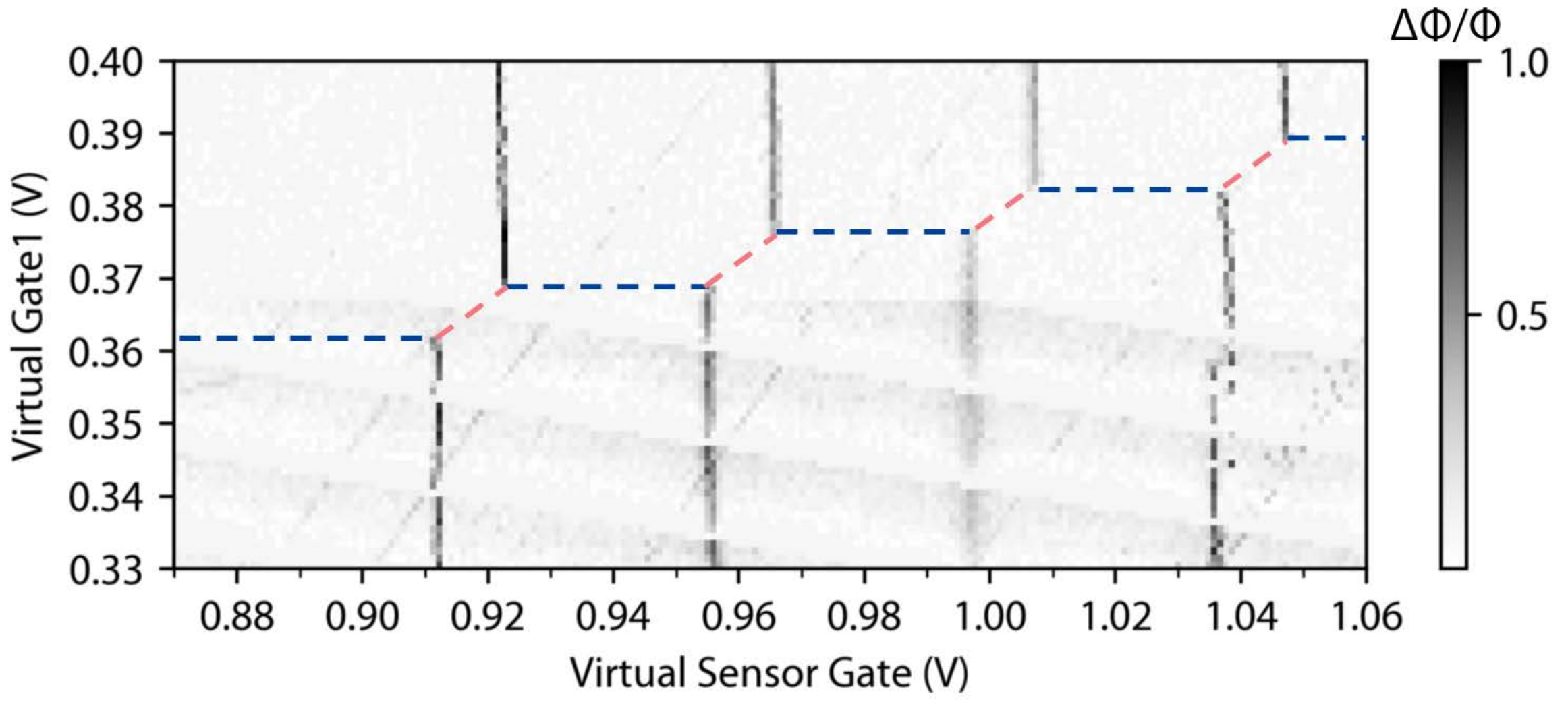}
\caption{Dot anticrossing produced in virtual gate space.} 
\label{SupFig:FigS2_VirtualGateMap}
\end{figure*}

Based on the diagonal elements in the above matrix, these simulated values for the lever arm are approximately $2/3$ of those extracted from experimental data. 
From here, we can transform into a virtual gate space following the methods in Ref.~\cite{volk2019}, where a change in the virtual voltage $\Delta\tilde{\bm{V}}$ is mapped to change in the physical input voltages $\Delta\bm{V}$ by means of the matrix $\bm{B}$, such that $\Delta\bm{V} = \bm{B} \cdot \Delta\tilde{\bm{V}}$. From the above capacitance matrix we find the following values for $\bm{B}^{\rm sim.}_{\rm 2x2}$, compared to the experimentally derived values $\bm{B}^{\rm exp.}_{\rm 2x2}$

\begin{equation}
\hspace*{-0.8cm}
\bm{B}^{\rm sim.}_{\rm 2x2}=
\begin{bmatrix}
1.0083  & -0.0219 & -0.0876 & -0.0110 \\
-0.0217 & 1.0083  & -0.0111 & -0.0875 \\
-0.0872 & -0.0110 & 1.0083  & -0.0217 \\
-0.0110 & -0.0875 & -0.0218 & 1.0083  \\
\end{bmatrix},\quad
\bm{B}^{\rm exp.}_{\rm 2x2}=
\begin{bmatrix}
 1.241 & -0.0745 & -0.943 &  0.207  \\
-0.265 &  1.1165  & 0.212 & -0.630 \\ 
-0.302 & -0.1234 &  1.418 & -0.396 \\ 
 0.090 & -0.1340 & -0.605 &  1.273 \\
\end{bmatrix}\label{SS_Eq:2x2Vitual}
\end{equation}
where the effect of the aforementioned underestimate of the simulations compared to experiments remains visible. 
As proof of principle for this method, we utilise the transformation matrix $\bm{B}^{\rm exp.}_{\rm 2x2}$ in order to produce the anti-crossing seen in Fig.~\ref{SupFig:FigS2_VirtualGateMap}.

Here dashed lines are added as a guide to the eye for the second dot-reservoir transition (blue) and inter-dot charge transitions (red), both which have tunnel rates outside the bandwidth of our reflectometry sensor. The virtual gate space allows us to easily depopulate the remaining two dots of any carriers, as is the case for this data. The diagonal background in this experiment is interference from a parallel-running setup on the same dilution refrigerator, with the abrupt disappearance of this periodic signal an indication of that experiment completing. 

\newpage
\section{Dual-nanowire 2$\times$4 array capacitance and lever arm matrix}\label{SupSec:2x4CapMat}
The study outlined above in \S\ref{SupSec:2x2CapMat} can be extended to include the floating gate device illustrated in Fig. 1 of the main text, as shown in Fig.~\ref{SupFig:FigS1_DeviceCompare}(b). Here it is important to note that the overlying metallic gate electrode, T, can have a strong tuning effect upon the floating gates. As such, we expand the analysis to include to the top gate, and the capacitance matrix between all elements of the $2\times4$ array of quantum dots is illustrated in Table~\ref{SS_Tab:2x4CapMat}. Each quantum dot is also expected to have additional small capacitances to the environment, such as source, drain, backgate. These capacitances are expected to be small and therefore ignored in our model.

From this capacitance matrix, we can write the lever-arm matrix as a function of the metallic electrodes in Eq.~\ref{SS_Eq:2x4LA_FG}. 
\begin{equation}
\hspace*{-0.5cm}\begin{pmatrix}
\mu_{\rm L1d}\\
\mu_{\rm L2d}\\
\mu_{\rm R1d}\\
\mu_{\rm R2d}\\
\mu_{\rm FL1d}\\
\mu_{\rm FL2d}\\
\mu_{\rm FR1d}\\
\mu_{\rm FR2d}\\
\end{pmatrix}
= e\times 10^{-2}
\begin{bmatrix}
13.2767 & 0.2743  & 0.0243  & 0.0125  & 0.6487 & 0.1175 & 0.0550 \\
0.2743  & 13.2524 & 0.0125  & 0.0243  & 0.1175 & 0.6487 & 0.0550 \\
0.0243  & 0.0125  & 13.2524 & 0.2743  & 0.6487 & 0.1176 & 0.0550 \\
0.0125  & 0.0242  & 0.2743  & 13.2282 & 0.1176 & 0.6487 & 0.0548 \\
0.5194  & 0.0956  & 0.0222  & 0.0108  & 5.5294 & 0.1139 & 0.0286 \\
0.0961  & 0.5218  & 0.0108  & 0.0223  & 0.1148 & 5.5798 & 0.0287 \\
0.0221  & 0.0107  & 0.5194  & 0.0954  & 5.5462 & 0.1141 & 0.0285 \\
0.0107  & 0.0221  & 0.0954  & 0.5194  & 0.1136 & 5.5126 & 0.0285 \\
\end{bmatrix}
\begin{pmatrix}
V_{\rm L1}\\
V_{\rm L2}\\
V_{\rm R1}\\
V_{\rm R2}\\
V_{\rm F1}\\
V_{\rm F2}\\
V_{\rm T}\\
\end{pmatrix}\label{SS_Eq:2x4LA_FG}
\end{equation}
which systematically falls under the same considerations discussed in Supplementary \S\ref{SupSec:2x2CapMat}.
Under the assumption that the input voltages are pinned to their respective supplies, we can utilise the fact that the voltage on the floating gate electrodes is determined by the capacitance matrix. As a first order approximation, we assume fully depleted dots, which results in the simplification of our virtual gate space to include only user controlled input voltages through the transformation:
\begin{equation}
\begin{pmatrix}
V_{\rm L1}\\
V_{\rm L2}\\
V_{\rm R1}\\
V_{\rm R2}\\
V_{\rm F1}\\
V_{\rm F2}\\
V_{\rm T}\\
\end{pmatrix}
=
\begin{bmatrix}
1 & 0 & 0 & 0 & 0\\
0 & 1 & 0 & 0 & 0\\
0 & 0 & 1 & 0 & 0\\
0 & 0 & 0 & 1 & 0\\
\alpha_{\rm ({L1},{F1})} & \alpha_{\rm ({L2},{F1})} & \alpha_{\rm ({R1},{F1})} & \alpha_{\rm ({R2},{F1})} & \alpha_{\rm ({T},{F1})}\\
\alpha_{\rm ({L1},{F2})} & \alpha_{\rm ({L2},{F2})} & \alpha_{\rm ({R1},{F2})} & \alpha_{\rm ({R2},{F2})} & \alpha_{\rm ({T},{F2})}\\
0 & 0 & 0 & 0 & 1\\
\end{bmatrix}
\begin{pmatrix}
V_{\rm L1}\\
V_{\rm L2}\\
V_{\rm R1}\\
V_{\rm R2}\\
V_{\rm T}\\
\end{pmatrix}\label{SS_Eq:2x4LA}
\end{equation}
where $\alpha_{(i,j)} = -C_{(i,j)}/C_{(j,j)}$ is the lever arm between electrodes. Hence, the transformation into a virtual gate space can be made here also, with the simulations yielding the control matrix $B^{\rm sim.}_{\rm 2x4}$:
\begin{equation}
\hspace*{-1.5cm}\begin{pmatrix}
\Delta V_{\rm L1}\\
\Delta V_{\rm L2}\\
\Delta V_{\rm R1}\\
\Delta V_{\rm R2}\\
\Delta V_{\rm T}\\
\end{pmatrix} = 
\begin{bmatrix}
1.4736  & 0.5051  & 0.4715  & 0.5244  & -1.2183 & -1.3752 & -1.2305 & -1.3845 \\
0.5006  & 1.4797  & 0.5195  & 0.4780  & -1.3628 & -1.2328 & -1.3429 & -1.2761 \\
0.4727  & 0.5253  & 1.4723  & 0.5041  & -1.2556 & -1.3800 & -1.1931 & -1.3794 \\
0.5188  & 0.4769  & 0.4973  & 1.4768  & -1.3626 & -1.2652 & -1.3330 & -1.2334 \\
-0.2813 & -0.2840 & -0.2807 & -0.2835 & 0.7187  & 0.7258  & 0.7058  & 0.7280   \\
\end{bmatrix}
\begin{pmatrix}
\Delta \tilde{V}_{\rm L1}\\
\Delta \tilde{V}_{\rm L2}\\
\Delta \tilde{V}_{\rm R1}\\
\Delta \tilde{V}_{\rm R2}\\
\Delta \tilde{V}_{\rm FL1}\\
\Delta \tilde{V}_{\rm FL2}\\
\Delta \tilde{V}_{\rm FR1}\\
\Delta \tilde{V}_{\rm FR2}\\
\end{pmatrix}
\end{equation}

It can be seen from this virtual gate space matrix, that the T top gate plays a pivotal role in the tunability of the the floating gates, and dots defined beneath with respect to the surrounding electrodes. 
\newpage
\KOMAoptions{paper=landscape}
\recalctypearea
\thispagestyle{empty}
\begin{table}[ht]

  \centering
    \addtolength{\leftskip} {-8cm} 
    \captionsetup{width=1.5\linewidth}
    \begin{tabular}{|c||c|c|c|c|c|c|c|c|c|c|c|c|c|c|c|c|c|}
    \hline
   & L1 & L2 & F1 & F2 & R1 & R2 & L1d & L2d & FL1d & FL2d & FR1d & FR2d & R1d & R2d  & T \\
  \hhline{|=||=|=|=|=|=|=|=|=|=|=|=|=|=|=|=|}
     L1& 41.9500 & -9.3764 & -9.4797 & -1.8504 & -0.3077 & -0.1311 & -5.6181 & -0.1301 & -0.1991 & -0.0349 & -0.0071 & -0.0033 & -0.0122 & -0.0059 & -5.9482\\ 
    \hline
   L2& -9.3764 & 41.9850 & -1.8479 & -9.4744 & -0.1312 & -0.3072 & -0.1294 & -5.6626 & -0.0341 & -0.2028 & -0.0033 & -0.0071 & -0.0059 & -0.0122 & -5.9464   \\
    \hline
   F1&-9.4796 & -1.8481 & 59.9050 & -14.9060 & -9.4092 & -1.8387 & -0.5071 & -0.0851 & -2.6947 & -0.0586 & -2.7288 & -0.0582 & -0.6052 & -0.0957 & -6.2944\\
    \hline
   F2&     -1.8506 & -9.4743 & -14.9060 & 59.9990 & -1.8397 & -9.4087 & -0.0845 & -0.5137 & -0.0572 & -2.7582 & -0.0577 & -2.7482 & -0.0961 & -0.6047 & -6.2999 \\
    \hline
    R1& 0.3076 & -0.1310 & -9.4092 & -1.8395 & 41.7510 & -9.3650 & -0.0108 & -0.0053 & -0.0070 & -0.0033 & -0.1969 & -0.0344 & -5.5215 & -0.1413 & -5.9494  \\
    \hline
    R2 &-0.1310 & -0.3070 & -1.8384 & -9.4087 & -9.3651 & 41.7780 & -0.0053 & -0.0109 & -0.0032 & -0.0071 & -0.0341 & -0.1978 & -0.1420 & -5.5487 & -5.9484 \\
    \hline
    L1d    & -5.6181 & -0.1295 & -0.5072 & -0.0844 & -0.0108 & -0.0053 & 6.7500 & -0.0086 & -0.0384 & -0.0029 & -0.0005 & -0.0003 & -0.0007 & -0.0004 & -0.0584 \\
    \hline
    L2d&    -0.1301 & -5.6625 & -0.0850 & -0.5138 & -0.0053 & -0.0110 & -0.0086 & 6.8084 & -0.0029 & -0.0397 & -0.0002 & -0.0005 & -0.0004 & -0.0007 & -0.0593 \\
    \hline
    FL1d&  -0.1990 & -0.0341 & -2.6947 & -0.0573 & -0.0070 & -0.0033 & -0.0384 & -0.0029 & 3.1842 & -0.0016 & -0.0005 & -0.0002 & -0.0005 & -0.0003 & -0.0227\\
    \hline
    FL2d&    -0.0349 & -0.2027 & -0.0586 & -2.7582 & -0.0033 & -0.0072 & -0.0029 & -0.0397 & -0.0016 & 3.2581 & -0.0002 & -0.0005 & -0.0003 & -0.0006 & -0.0231  \\
    \hline
    FR1d&-0.0071 & -0.0033 & -2.7287 & -0.0576 & -0.1968 & -0.0341 & -0.0005 & -0.0003 & -0.0004 & -0.0002 & 3.2270 & -0.0016 & -0.0470 & -0.0032 & -0.0230  \\
    \hline
    FR2d& -0.0033 & -0.0072 & -0.0582 & -2.7482 & -0.0344 & -0.1978 & -0.0003 & -0.0005 & -0.0002 & -0.0005 & -0.0016 & 3.2493 & -0.0033 & -0.0472 & -0.0231  \\
    \hline
    R1d&-0.0122 & -0.0059 & -0.6053 & -0.0960 & -5.5214 & -0.1420 & -0.0007 & -0.0004 & -0.0005 & -0.0003 & -0.0470 & -0.0033 & 6.8241 & -0.0104 & -0.0653 \\
    \hline
    R2d&-0.0059 & -0.0122 & -0.0956 & -0.6048 & -0.1413 & -5.5486 & -0.0004 & -0.0007 & -0.0003 & -0.0005 & -0.0032 & -0.0472 & -0.0104 & 6.8509 & -0.0656 \\
    \hline
    T& -5.9489 & -5.9471 & -6.2951 & -6.3006 & -5.9480 & -5.9471 & -0.0586 & -0.0594 & -0.0227 & -0.0231 & -0.0230 & -0.0230 & -0.0651 & -0.0654 & 100.3100 \\
    \hline
    \end{tabular}%
\caption{Maxwell capacitance matrix (in aF) for the $2\times4$ array of quantum dots distributed across two parallel nanowires with interconnecting floating gates and overlying metallic top gate.}
    \label{SS_Tab:2x4CapMat}
\end{table}%
\clearpage
\KOMAoptions{paper=portrait}
\recalctypearea

\newpage
\section{Floating gate Capacitive effects : Extended analysis}\label{SupSec:FloatCapEffects}
\begin{figure*}[ht]
\centering
\includegraphics[width=1\linewidth]{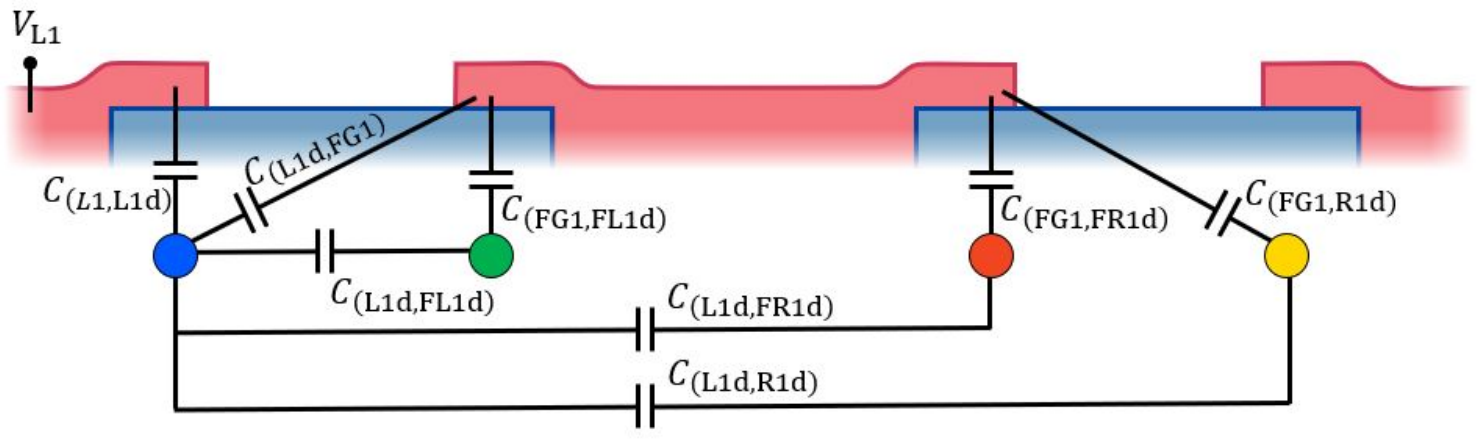}
\caption{Simplified capacitive network of the dual-nanowire device illustrating a 1$\times$4 array slice. The influence of the floating gate electrode is captured by the additional cross-capacitances highlighted.} 
\label{SupFig:CapNetwork}
\end{figure*}

In order to analyse the effect of the floating gate electrode, we take into account the second order shift in the chemical potential of the sensor dot L1d of the form dot~$\rightarrow$~FG~$\rightarrow$~sensor. With respect to the simplified device capacitance network as shown in Fig.~\ref{SupFig:CapNetwork}, we treat the floating gate in a similar fashion to the quantum dots from Ref.~\cite{vanderwiel2003a}, while maintaining a fixed charge to reflect the electrical isolation of the gate. 

\begin{figure*}[ht]
\centering
\includegraphics[width=1\linewidth]{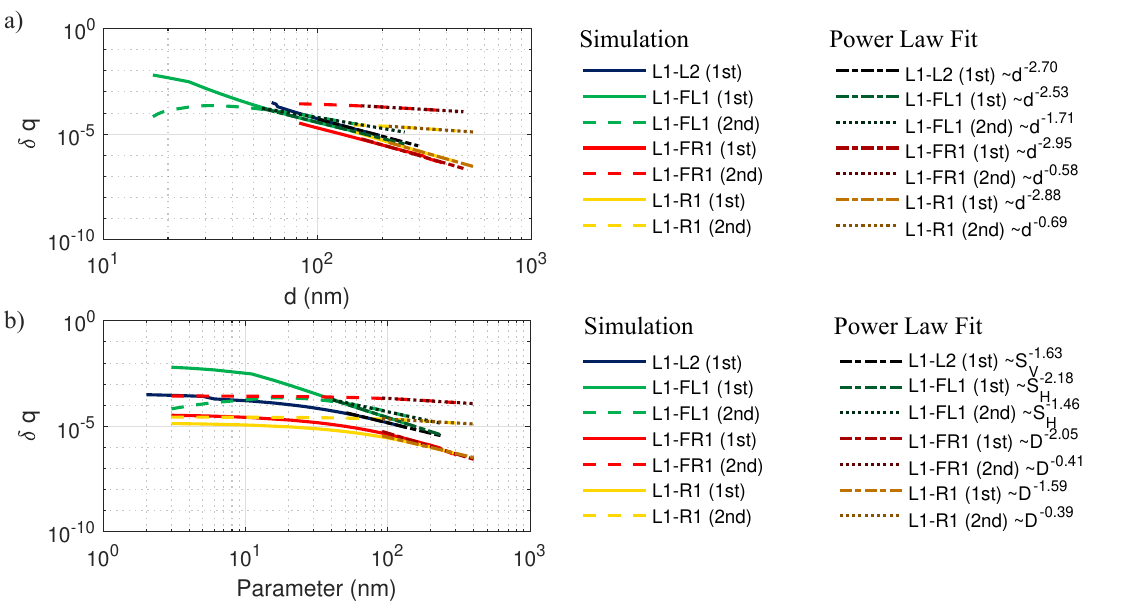}
\caption{(a) Illustration of the first order and second order solutions to the detected charge on the sensor, normalised to sensor dot charge, as a function of the dot-dot separation. (b) Illustrates the decay rate as a function of the design elements within the device architecture.}
\label{SupFig:OrderApprox}
\vspace{0cm}
\end{figure*}

Following the analysis in Ref.~\cite{vanderwiel2003a}, a ``first order'' sensor voltage shift, which is due to the addition of an electron and direct dot-dot mutual capacitances is given by:

\begin{align}
    \Delta V_{\rm (L1,FL1d)}^{\rm (1)} &= \frac{|e|}{C_{(\rm L1, L1d)}}\cdot \frac{C_{\rm (L1d,FL1d)}}{C_{\Sigma \rm FL1d}}\\
    \Delta V_{\rm (L1,FR1d)}^{\rm (1)} &= \frac{|e|}{C_{(\rm L1, L1d)}}\cdot \frac{C_{\rm (L1d,FR1d)}}{C_{\Sigma \rm FR1d}}\\
    \Delta V_{\rm (L1,R1d)}^{\rm (1)} &= \frac{|e|}{C_{(\rm L1, L1d)}}\cdot \frac{C_{\rm (L1d,R1d)}}{C_{\Sigma \rm R1d}}
\end{align}
where the elements can be extracted from the Maxwell capacitance matrix, detailed in \S\ref{SupSec:2x4CapMat}, for each data-point in the parametric sweeps described in the main text. From Fig.~\ref{SupFig:OrderApprox}, it can be seen that this first order effect is highly suppressed where there is a large separation between the dots, as the direct mutual capacitance between the dots and the SEB sensor L1d rolls off with distance with a power law ranging from $\Delta q \propto d^{-3.0}$ to $d^{-2.5}$ (additional decay fits can be seen in Fig.~\ref{SupFig:OrderApprox}). Most of these decays fall close to the $d^{-3}$ dependence measured for planar devices in silicon~\cite{Zajac2016,neyens2019}, possessing a high density of metallic electrodes present that can contribute to a screening effect of the charge. This is contrasted with the face-to-face decay rate of $d^{-2.5}$, where the majority of the metal between the two dots has been removed. If we now take the effective shift in the floating gate into account we arrive at a ``second order'' approximation to the sensor voltage shift:
\begin{align}
    \Delta V_{\rm (L1,FL1d)}^{\rm (2)} &= \frac{|e|}{C_{(\rm L1, L1d)}} \left[\frac{C_{\rm (L1d,FL1d)}}{C_{\Sigma \rm FL1d}}+\frac{C_{\rm (L1d,FG1)}}{C_{\Sigma \rm FG1}}\cdot\frac{C_{\rm (FG1,FL1d)}}{C_{\Sigma \rm FL1d}}\right]\\
    \Delta V_{\rm (L1,FR1d)}^{\rm (2)} &= \frac{|e|}{C_{(\rm L1, L1d)}}\left[ \frac{C_{\rm (L1d,FR1d)}}{C_{\Sigma \rm FR1d}}+\frac{C_{\rm (L1d,FG1)}}{C_{\Sigma \rm FG1}}\cdot\frac{C_{\rm (FG1,FR1d)}}{C_{\Sigma \rm FR1d}}\right]\\
    \Delta V_{\rm (L1,R1d)}^{\rm (2)} &= \frac{|e|}{C_{(\rm L1,L1d)}}\left[ \frac{C_{\rm (L1d,R1d)}}{C_{\Sigma \rm R1d}}+\frac{C_{\rm (L1d,FG1)}}{C_{\Sigma \rm FG1}}\cdot\frac{C_{\rm (FG1,R1d)}}{C_{\Sigma \rm R1d}}\right]
\end{align}
where the $C_{\rm (L1d,FG1)}$ term represents the coupling between the sensor and the floating gate, and $C_{({\rm FG1},i)}/C_{\Sigma i}$ represents the charge capacitively induced on the floating gate, distributed by a factor of $1/C_{\Sigma \rm FG1}$. It is the total capacitance of the floating gate which is then subject to geometrical dependencies upon the parameter sweeps. The comparisons within Fig.~\ref{SupFig:OrderApprox}(a) illustrates that, for sensing dots in the remote nanowire, the second-order contribution due to the floating gate is dominant, giving rise to the advantage of the floating gate electrodes for long-range capacitive sensing.
While the core aim of this manuscript is to compare the additional sensitivity to charge movements in remotely located quantum dots as facilitated by the floating gate, for completeness we note this second order effect of the floating gate also enhances to sensitivity to FL1d (located within the same nanowire as the sensor, but under the floating gate). Such an enhancement is absent for the equivalent dot in a single-nanowire 2$\times$2 array, where all electrodes are pinned to a supply voltage, and the trend lines  shown in the main text for face-to-face dots in the same nanowire therefore consider only this first-order effect. 

While effects of dot separation on mutual capacitance has been well studied in the literature~\cite{Zajac2016,neyens2019}, we wish to quantify the effects of the specific design parameters of these nanowire QD devices on the mutual capacitance. As shown in Fig.~\ref{SupFig:OrderApprox}(b) we have re-scaled the $x$-axis so that it is with respect to the input design parameters described in Fig.~\ref{SupFig:FigS1_DeviceCompare}. Here we can see that the trends plateau as each design parameter begins to approach the size of the quantum dot, moving to a regime which would be physically challenging to realise in fabrication. The reduction in sensitivity due to the input parameter follows a different trend when compared to the dot-dot distance $d$. The reduction in sensitivity due to the face-to-face gap between electrodes $S_{\rm h}$ is the largest, at $\Delta q\propto S_{\rm h}^{-2.18}$, reduced to $\Delta q\propto S_{\rm h}^{-1.46}$ when this dot is located beneath a floating gate. The $S_{\rm v}$ gap between the two gates along the nanowire gives rise to $\Delta q\propto S_{\rm v}^{-1.63}$. For the floating gate, the dimensions of the floating gate contribute to the self-capacitance of the electrode which, in turn, will contribute to the decay in sensitivity. For the floating gate geometry in this work, we observe a $\Delta q\propto D^{-0.41}$, however, this could be subject to further device optimisation, outside the scope of this work, but discussed in the context of GaAs planar devices in Ref.~\cite{trifunovic2012a}.

\newpage

\section{Estimates for qubit-qubit coupling: Extended analysis}\label{SupSec:Estimates2Q}

A number of possible qubit-qubit coupling mechanisms were identified in the main text, which utilise the floating gate to couple two qubits situated within different nanowires. These estimates are born from analysis in the wider literature, but discussed in the context of the silicon MOS devices here. Namely, a focus on the analysis presented in Ref.~[\citen{trifunovic2012a}] and Ref.~[\citen{neyens2019}] is discussed.

\begin{figure*}[ht]
\centering
\includegraphics[width=1\linewidth]{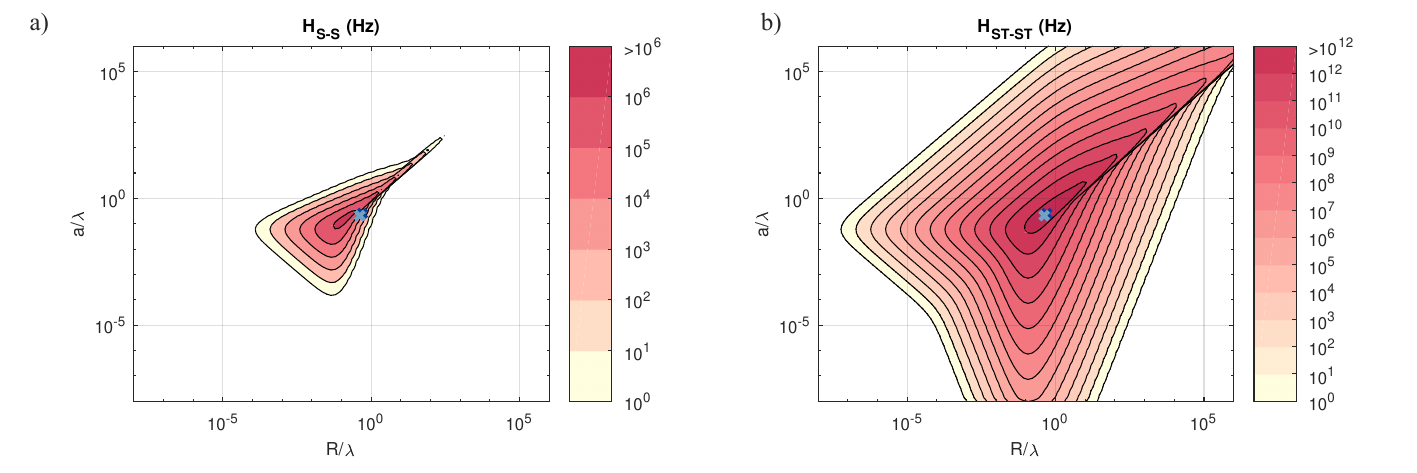}
\caption{Contour maps demonstrating the predicted coupling strength for qubit-qubit interactions $H_{\rm S-S}$, between two electron spin qubits and $H_{\rm ST-ST}$, between two Singlet-Triplet qubits. These are presented as the unitless ratio between core dimension parameters, where $\lambda$ is the dot length scale, $R$ is the effective radius of the electrode and $a$ is the offset between the effective centre of the electrode and the centre of the dot. The two devices studied in this work are shown as the overlapping blue markers.
}
\label{SupFig:2QEst}
\end{figure*}

The data seen in Fig.~\ref{SupFig:2QEst} illustrates the solution to the maximal spin-spin coupling strength $H_{\rm S-S}$ shown in Eq.~(9) of Ref.~[\citen{trifunovic2012a}] as well as the estimate described in Ref.~[\citen{trifunovic2012a}] for $H_{\rm ST-ST}$ via an electric dipole created across the nanowire. Here, the critical length scales include the effective radius of the electrode above the quantum dot defined as $R$, the dot length scale $\lambda$ and $a$, the offset between the effective centre of the electrode and centre of the dot.

The simulations in Ref.~[\citen{trifunovic2012a}] focus on deriving qubit-qubit coupling strengths for GaAs and InAs quantum dots, where the ratio $\lambda/\lambda_{SO}$ is $\sim 10^{-1} - 1$, with the effective spin-orbit length $\lambda_{SO} = \hbar/(m^*\alpha)$ and $\alpha$ as the Rashba spin-orbit interaction strength. A critical difference between these materials and silicon is the magnitude of the measured $\alpha \sim 15.2\times 10^{-13}$eVcm for a silicon quantum dot device~\cite{tanttu2019} placing $\lambda/\lambda_{SO} \sim 10^{-4}$ and yielding a weaker $H_{\rm S-S}$. Conversely, the solution for $H_{\rm ST-ST}$ does not depend on $\lambda_{SO}$, but instead upon device geometry, and in fact the silicon device geometry interestingly becomes more favourable due to reductions in the distance $t_{\rm ox}$ between the electrodes and the quantum dot as discussed in Ref.~[\citen{trifunovic2012a}].

Considering the interactions between two charge qubits through the floating gate, we refer to the discussions in Ref.~[\citen{neyens2019}] which couples in a similar form when compared to the $H_{\rm ST-ST}$ qubits. We simply utilise the formula below which is the equivalent of Eq.~(S20) in Ref.~[\citen{neyens2019}]:

\begin{equation}
    g = e^2 \dfrac{\left(\splitdfrac{
    C_{\Sigma L1d} C_{\Sigma R1d} C'_{FL1d,FR1d} 
    - C_{\Sigma L1d} C'_{FL1d,FR1d} C_{FR1d,R1d}
    }{- C_{\Sigma R1d} C_{L1d,FL1d} C'_{FL1d,FR1d}
    -C_{L1d,FL1d} C'_{FL1d,FR1d} C_{FR1d,R1d}
    }\right)}{\left(\splitdfrac{
    C_{\Sigma L1d}C_{\Sigma FL1d}C_{\Sigma FR1d}C_{\Sigma R1d}
    - C_{\Sigma L1d}C_{\Sigma FL1d}C_{FR1d,R1d}^2
    }{- C_{\Sigma L1d}C_{\Sigma R1d}C_{FL1d,FR1d}^{'2}
    - C_{\Sigma FR1d}C_{\Sigma R1d}C_{L1d,FL1d}^2
    - C_{L1d,FL1d}^2 C_{FR1d,R1d}^2
    }\right)},
\end{equation}
where the equivalent value for $C'_{FL1d,FR1d}$ is enhanced via the floating gate electrode as discussed in earlier sections, and the values for each capacitance is derived from our Maxwell capacitance matrix in Table~\ref{SS_Tab:2x4CapMat}.

\newpage 

\bibliographystyle{achemso}

\providecommand{\latin}[1]{#1}
\makeatletter
\providecommand{\doi}
  {\begingroup\let\do\@makeother\dospecials
  \catcode`\{=1 \catcode`\}=2 \doi@aux}
\providecommand{\doi@aux}[1]{\endgroup\texttt{#1}}
\makeatother
\providecommand*\mcitethebibliography{\thebibliography}
\csname @ifundefined\endcsname{endmcitethebibliography}
  {\let\endmcitethebibliography\endthebibliography}{}